\documentclass[aps,prd,onecolumn,showpacs,showkeys,superscriptaddress]{revtex4-2}
\usepackage{latexsym}
\usepackage{amssymb}
\usepackage{amsmath}
\usepackage{amscd}
\usepackage{amsthm}
\usepackage{graphicx}
\usepackage{textcomp}
\usepackage{colortbl}
\usepackage[colorlinks]{hyperref}
\usepackage[font={footnotesize,it}]{caption}
\usepackage{multirow}
\usepackage[T1]{fontenc}
\usepackage{ae,aecompl}
\usepackage{subcaption}
\begin{document}
\title{Observational Constraints on generalized dark matter properties in the presence of neutrinos with the final Planck release}
\author{Vikrant Yadav}
\email{vikuyd@gmail.com}
\affiliation{School of Basic and Applied Sciences, Raffles University, Neemrana - 301705, Rajasthan, India}

\author{Santosh Kumar Yadav}
\email[]{sky91bbaulko@gmail.com}
\affiliation{School of CS \& AI, SR University, Warangal - 506002, Telangana, India.}
\author{Anil Kumar Yadav}
\email[]{abanilyadav@yahoo.co.in}
\affiliation{Department of Physics, United College of Engineering and Research,Greater Noida - 201310, India}

\begin{abstract}
In this paper, we investigate an extension of the standard $\Lambda$CDM model by allowing: a temporal evolution in the equation of state (EoS) of DM via Chevallier-Polarski-Linder parametrization, and the constant non-null sound speed. We also consider the properties of neutrinos, such as the effective neutrino mass and the effective number of neutrino species as free parameters.  We derive the constraints on this scenario by using the data from the Planck-2018 cosmic microwave background (CMB), baryonic acoustic oscillation (BAO) measurements, Pantheon+ compilation of Type Ia supernovae (SNe Ia), and some large scale structure (LSS) information from the cosmic shear surveys: Kilo Degree Survey (KiDS)-1000 and Dark Energy Survey (DES). We find constraints on the EoS and sound speed of DM very close to the null value in all the analyses, and thus no significant evidence is found beyond the standard CDM paradigm. In all the analyses, we find the significantly tight upper bounds on the sum of neutrino masses, and significantly lower mean values of $S_8$, which are in agreement with the LSS measurements. Thus, the well-known $S_8$ tension is reconciled in the considered model.
\end{abstract}


\maketitle
\section{Introduction}
\label{sec:intro}
The well-known concordance $\Lambda$CDM model provides a good fit to the various cosmological observations of different physical origins. This six-parameter cosmological model represents the minimal and simplest framework to provides all available cosmological information. One of the key ingredients of this standard model is dark matter (DM), necessary for the formation of observed structures in the Universe. DM is thought to contribute about 26\% to the total energy density of the Universe today \cite{Planck2015, Planck2018} but its particle nature is still mysterious. Many efforts have been made to detect DM candidate via direct and indirect searches but no concrete candidate is found so far. Although there are a plethora of physically motivated candidates of DM, all are assumed to behave as a pressureless perfect fluid. For more details about the pieces of evidence, candidates and methods of detection of DM, see the review articles \cite{DM01, DM02, DM03}, and for the current status of direct and indirect searches of DM particles, see \cite{DM04, DM05}. One may see the recent article on  review of the particle physics, astrophysics, and cosmology associated with the cosmological tensions and anomalies in \cite{snowmass1}.\\

In the standard model, the modelling of DM as a pressureless perfect fluid, usually dubbed as cold dark matter (CDM) leads to many small scale issues which can be seen in the articles \cite{SSP01,SSP02, SSP03,SSP04}. Therefore, it is important to place the constraints on different quantities which are helpful to describe/characterize the physical nature of DM, with the latest cosmological observations. One of the key quantity which plays an important role in modern cosmology is the equation of state (EoS) parameter of DM and its value should be fixed by observations. The EoS parameter of CDM is zero but there is no preferred reason for it to be zero. In fact, the precise nature of DM could be identified by more general parameters such as a time-dependent EoS and a non-zero sound speed. In literature, many attempts have been made to know the particle nature of DM  by constraining its EoS as a constant, time-variable and/or a more general mathematical form, with the observational data sets. 
See a partial list \cite{wdm00,wdm01,wdm02,wdm03,wdm04,wdm05,wdm06,wdm07,wdm08,wdm09} of works in this direction, and also recent investigations   \cite{gdm00,gdm01,gdm02,gdm04,mnras, ED22}.\\

The present day expansion rate of the Universe is described by the Hubble constant $H_0$, and the measurements of the r.m.s. fluctuation of density perturbation at $8h^{-1}\, {\rm Mpc}$ are quantified by the parameter $\sigma_8$ or $S_8\equiv\sigma_8\sqrt{\Omega_{\rm m}/0.3}$, where $\Omega_{\rm m}$ is the present day matter density parameter. In recent years in the era of precision cosmology,  statistically significant tensions have emerged in the parameters $H_0$ and $S_8$. There is around 5$\sigma$ tension in $H_0$, which refers to the difference between its direct local distance ladder measurements, viz.,  the SH0ES measurement $H_0 = 73.04\pm 1.04{\rm \,km\, s^{-1}\, Mpc^{-1}}$ (68\% CL)~\cite{Riess:2021jrx} based on the Type Ia supernovae (SNe Ia) calibrated by Cepheid variables, and $H_0 = 67.27\pm 0.60{\rm \,km\, s^{-1}\, Mpc^{-1}}$ (68\% CL)~\cite{Planck2018} inferred from \textit{Planck} 2018 assuming the standard $\Lambda$CDM model. Likewise, there is around $3\sigma$ tension in $S_8$ within the framework of $\Lambda$CDM, which refers to the difference in its estimations, $S_8=0.834 \pm 0.016$ inferred from \textit{Planck} 2018 ~\cite{Planck2018}, and from cosmic shear surveys eg. Kilo-Degree Survey (KiDS-1000) $S_8=0.759 \pm 0.022$ \cite{KiDS:2020suj} and Dark Energy Survey (DES) $S_8=0.776 \pm 0.017$ \cite{DES:2021wwk}). Extended $\Lambda$CDM scenarios are extensively explored in the literature in the context of these two tensions \cite{Abdalla,Perivolaropoulos:2021jda,Kumar:2016zpg,Kumar:2019wfs,Kumar:2021eev,deAraujo:2021cnd,Hu:2023jqc,Vagnozzi:2023nrq,Kroupa:2023ubo,Bernui:2023byc,Akarsu:2021fol,Akarsu:2022typ,Akarsu:2023mfb}.\\

The neutrinos play an important role in the background evolution as well as in the structure formation with direct implications on Cosmic Microwave Background (CMB), Large Scale Structure (LSS) and nuclear abundances produced in the Big Bang Nucleosynthesis (BBN) epoch, see the recent investigations with the neutrino properties in \cite{Lattanzi,Giusarma,Nunes,kumar,Choudhury1, Choudhury2, Valentino, vagnozi22, Haili22,Vagnozzi/2017,Vagnozzi/2018,Giusarma/2018} and references therein, and \cite{N1,N2,N3} for a review. As long as neutrinos were relativistic, they contribute to radiation density and the effect is paramerterized by the term, $N{\rm eff}$. However, at late times, when they turn to non-relativistic, they contribute to total matter density. It is also known today from the results on flavour neutrino oscillation experiments that neutrinos are massive. At least one neutrino state has a large enough mass for being non-relativistic today, thus making up a small fraction of the dark matter of the Universe. The oscillation experiments are able to measure two of the neutrino mass-splittings and also provide the information of neutrino-mass ordering \cite{salas}. Although, the neutrino oscillation experiments do not provide any hint about the absolute scale of neutrino masses. The bounds on neutrino mass scale can be placed by the cosmological observations and at the moment cosmology provides a stronger upper bound on the sum of neutrino masses \cite{Planck2018}. But the challenge with the cosmological observations is to go beyond a tight upper bound. The absolute scale of neutrino mass is also placed by laboratory searches such as beta-decay \cite{beta1} and neutrino-less double beta decay \cite{beta2,beta3}. Nowadays, with the accumulation of a large number of cosmological observations, cosmology has reached in a high precision era where it is unavoidable to take into account the presence of neutrino properties in the extended cosmological investigations. Also, such investigations are necessary in order to place new more accurate constraints on the full baseline of the non-standard cosmological models.\\

Extra relativistic degrees of freedom at recombination, parametrized by the number of equivalent light neutrino species $N_{\rm eff}$ Steigman et al \cite{Steigman/1977}, gives the most simplest extension of $\Lambda$CDM. In Refs. \cite{Akita/2020,Benetti/2021}, the authors have investigated $N_{\rm eff}^{\rm SM} = 3.044$ for massless neutrino families. In this paper, we are interested in constraining the generalized DM parameters: a time-dependent EoS parameter of DM via well-motivated Chevallier-Polarski-Linder (CPL) parameterization and the non-null sound speed of DM,  in the presence of neutrino properties.  We constrain this scenario with  the observational data from different sources, and explore whether the considered scenario leads to the relaxation of $H_0$ and $S_8$ tesnsions. The paper is structured as follows: In the next section, we present the cosmological model with extended DM properties. The methodology and the data sets used to constrain the model parameters are described in section \ref{data}. In section \ref{results}, we derive the observational constraints and discuss the results in detail. The concluding remarks of the study are presented in section \ref{CR}.

\section{Model with extended properties of dark matter and neutrinos}
\label{the_model}
In the present study, We consider the spatially-flat, homogeneous and isotropic  Friedmann-Lema\^{i}tre-Robertson-Walker (FLRW) Universe, given by the metric
\begin{equation}
    ds^2 = -dt^2 + a^2(t)\bigg[dr^2 + r^2 (d\theta^2 + \sin^2 \theta d\phi^2)\bigg],
\end{equation}
where we have adopted speed of light $c=1$; $a(t)$ denotes the cosmic scale factor of the Universe,  and $t$ is the cosmic time. Considering the above metric, the basic equations which determine the background evolution of the Universe, that is, the so-called Friedmann equations are given by
\begin{align}\label{model}
3 H^2 &=  8 \pi G (\rho_{\gamma } + \rho_{\nu}+ \rho_{\rm b} + \rho_{\rm dm} +\rho_{\Lambda}),\\ 
2 \frac{dH}{dt} + 3 H^2 &= - 8 \pi G (p_{\gamma} + p_{\nu}+ p_{\rm b} + p_{\rm dm} + p_{\Lambda}),
\end{align}
where $G$ is the Gravitational Newtonian constant and  $H = \frac{da/dt}{a}$ is the Hubble parameter which describes expansion rate of the Universe. In the above equations, $\rho$ and $p$ with subscripts denote the density and pressure of the different energy components of the Universe, namely,  photons ($\gamma$), neutrinos ($\nu$), baryons ($\rm b$), dark matter ($\rm dm$), and cosmological constant ($\Lambda$), respectively.

Under the assumption of no mutual interaction (except the usual gravitational interaction) between the above-mentioned species, the energy conservation equation reads as
\begin{align} \label{continiuty}
    \dot{\rho}_i + 3 H (1+w_i)\rho_i=0,
\end{align}
where $ i \in \{ \gamma,\nu,\rm b, \rm dm, \Lambda\}$ and $w_i$ is the EoS of $i \rm th$ species with $p_i =w_i \rho_i$. From eq. (\ref{continiuty}), it is easy to write energy density evolution of $i \rm th$ species with known EoS parameter, $w_i$.

In this article, we are not modelling DM as a pressureless perfect fluid. Rather, we assume a temporal evolution in the EoS of DM with the cosmic expansion. In order to quantify the temporal evolution with the cosmic evolution, we consider the well-motivated CPL  parametrization \cite{CPL01,CPL02} of EoS of DM, given by
\begin{align}\label{CPL}
w_{\rm dm}(a) &= w_{\rm dm 0} + w_{\rm dm 1}(1-a),
\end{align}
where $w_{\rm dm0}$ and  $w_{\rm dm1}$ are free parameters (constants) to be fixed by observations. The quantity $w_{\rm dm 0}$ represents the present value of EoS of DM. 
From eqs. (\ref{continiuty}) and (\ref{CPL}), the energy density evolution of DM is given by
\begin{align}
\rho_{\rm dm} &= \rho_{\rm dm0} a^{-3(1+w_{\rm dm0} + w_{\rm dm1})}\, e^{-3 w_{\rm dm1}(1-a)}.
\end{align}

At perturbative level, we choose to work in the conformal Newtonian gauge where the perturbed FLRW metric takes the following form
\begin{eqnarray}
\begin{aligned}
 ds^2  = a^2(\tau) \Big[-(1+2\psi) d\tau^2 + (1 - 2\phi)d\vec{r}^2 \Big],
\end{aligned}
\end{eqnarray}
where $\phi$ and $\psi$ are the metric potentials and $d\vec{r}$ represents the three spatial coordinates. Following the  notations of \cite{MB} and assuming small perturbations in the first order of $\nabla_\mu T^{\mu \nu}_{\rm dm} = 0$ in the Fourier space, we find the equations \cite{mnras}  
\begin{eqnarray}
\begin{aligned}
 \dot{\delta}_{\rm dm}=-(1+w_{\rm dm}) \left(\theta_{\rm dm} - 3 \dot{\phi} \right)   
 -3 \mathcal{H} \delta_{\rm dm} (\hat{c}^2_{\rm s,dm} - w_{\rm dm}) - 9 (1+w_{\rm dm})(\hat{c}^2_{\rm s,dm} - c^2_{\rm a,dm})\mathcal{H}^2 \frac{\theta_{\rm dm}}{k^2},
\end{aligned}
\end{eqnarray}
\begin{eqnarray}
\begin{aligned}
\dot{\theta}_{\rm dm}&=&-(1-3 \hat{c}^2_{\rm s,dm}) \mathcal{H} \theta_{\rm dm}  + \frac{\hat{c}^2_{\rm s,dm}}{1+w_{\rm dm}}k^2 \delta_{\rm dm} + k^2\psi. 
\end{aligned}
\end{eqnarray}
Here, the dot over a quantity denotes conformal time derivative, $\mathcal{H}$ is the conformal Hubble parameter and $k$ is magnitude of the wavevector $\vec{k}$. Further, $\delta_{\rm dm} = \delta \rho_{\rm dm} / \rho_{\rm dm}$ and $(\rho_{\rm dm} + p_{\rm dm})\theta_{\rm dm} = i k^j \delta T^{0}_{j}$  are the density perturbation and velocity perturbation, respectively, associated with DM fluid. The sound speed of DM in the rest frame is denoted by $\hat{c}^2_{\rm s,dm}$ and adiabatic squared sound speed is denoted by $c^2_{\rm a,dm}$.  
We have taken $\hat{c}^2_{\rm s,dm}$ as a free model parameter in order to get an insight of the micro-scale property of DM. If it significantly deviates from zero in light of the recent cosmological observations, it would be an indication for the DM  to be more complicated than simple CDM.

With the above given background and perturbation evolution equations, we present the cosmological model which is the extension of the standard $\Lambda$CDM model via the extended DM parameters plus the neutrino properties.
We consider the presence of three active neutrino species with the sum of total masses $\sum m_\nu$, where the mass ordering of the active neutrinos is fixed to the normal hierarchy, which is favored over the inverted one with current observations, as argued in \cite{huang,yang,guo}. Also, we take  $N_{\rm eff}$ as a free parameter to be fixed by observations, in order to verify any possible new correlation with the DM parameters ($w_{\rm dm0}, \, w_{\rm dm1}, \, \hat{c}^2_{\rm s,dm}$). We refer this scenario by $\Lambda_\nu$WDM and the full baseline parameters set is given by
 \begin{equation}
 \label{baseline2}
  \begin{aligned}
   \mathcal{P}_{\Lambda_{\nu} \rm WDM}=\Big\{\omega_{\rm b}, \, \omega_{\rm dm}, \, \theta_s, \, A_s, \, n_s,  
   \,\tau_{\rm reio},  \, w_{\rm dm0},
   \, w_{\rm dm1}, \, \hat{c}^2_{\rm s,dm},\, N_{\rm eff}, \, \sum m_{\nu} \Big\}.
 \end{aligned}
 \end{equation}
where the first six parameter correspond to $\Lambda$CDM and have their usual meaning \cite{Planck2015} whereas the remaining five are additional model parameters of our main interest. \\

\section{Data sets and  methodology}
\label{data}
We use the following data sets to derive observational constraints on the model parameters under consideration. \\

\noindent\textbf{CMB}: 
We consider the cosmic microwave background measurements from the Planck-2018 final release \cite{Planck2018,aghanim2,aghanim3}. We use high-$l$ and low-$l$ CMB  TT likelihood in the multipole ranges $30\leq l \leq 2508$ and $2\leq l \leq 29$, respectively,  along with the high-$l$ E mode polarization and temperature-polarization cross correlation likelihood.  We use low-$l$ E mode polarization data and the power spectra of the lensing potential. We refer this data simply as CMB throughout the manuscript.\\

\noindent\textbf{BAO}: 
The final Baryon Acoustic Oscillation (BAO) measurements of the SDSS collaboration cover eight different redshift intervals, and include anisotropic BAO measurements of $D_M(z)/r_d$ and $D_H(z)/r_d$ (where $D_M(z)$ is the comoving angular diameter distance and $D_H(z)=c/H(z)$ is the Hubble distance). Table 3 of Ref. \cite{eBOSS:2020yzd} compiles all the aforementioned BAO-only measurements.\\


\noindent\textbf{PP}: 
 We utilize distance modulus measurements of Type Ia supernovae (SNe Ia) obtained from the Pantheon+ sample, as described in \cite{Scolnic/2022}. This dataset includes 1701 light curves corresponding to 1550 different SNe Ia events distributed across the redshift range $z \in [0.001, 2.26]$. The dataset is conveniently referred to as Pantheon Plus (PP).\\

\noindent\textbf{KiDS-1000 and DES}:
 We use Gaussian priors of $S_8$ obtained by Kilo-Degree Survey (KiDS-1000) ($S_8=0.759 \pm 0.022$, \cite{KiDS:2020suj}) and Dark Energy Survey (DES) ($S_8=0.776 \pm 0.017$, \cite{DES:2021wwk}). \\

We have implemented the model in publicly available \texttt{CLASS} \cite{class} code interfaced with parameter inference code \texttt{MontePython} \cite{monte} which is embedded with  Metropolis-Hastings algorithm. We have used the uniform priors displayed in Table \ref{tab:priors} on all free model parameters to obtain correlated  Monte Carlo Markov Chain samples. The constraints on the model parameters are otained with five different sets of data combinations: CMB, CMB+BAO, CMB+BAO+PP, CMB+BAO+PP+KiDS-1000 and CMB+BAO+PP+DES.  The convergence of the Monte Carlo Markov Chains for all the model parameters have been ensured with Gelman-Rubin criterion \cite{Gelman_Rubin}. We have used the GetDist Python package \cite{antonygetdist} to analyze the samples. 



\begin{table}
\caption{Uniform priors on the parameters of $\Lambda_\nu$WDM model.} \label{tab:priors}
\begin{center}
\begin{tabular}{c c}
\hline \hline
Parameter & Prior\\
\hline
$100 \omega_{\rm b}$ & [0.8, 2.4]\\
$\omega_{\rm dm}$ & [0.01, 0.99] \\
$100\theta_s$ & [0.5, 2.0] \\
$\ln[10^{10}A_{s }]$ & [2.7, 4.0]\\
$n_s$ & [0.9, 1.1] \\
$\tau_{\rm reio}$  & [0.01,  0.8] \\
$w_{\rm dm0}$  & [0, 0.1] \\
$w_{\rm dm1}$ &  [0, 0.1] \\
$\hat{c}^2_{\rm s,dm}$ & [0, 0.1]\\
$\sum m_{\rm \nu}$ & [0.06, 1.0]\\
$N_{\rm eff}$ & [1,  4]\\
\hline \hline
\end{tabular}
\end{center}
\end{table} 

\section{Results and discussion}
\label{results}

\begin{table*}[!ht] 
\caption{\label{Table_M2} {Constraints with mean value and 68\% CL errors  on the free parameters and some derived parameters of $\Lambda_{\nu}$WDM model for five different data combinations, except the parameters $w_{\rm{dm0}}$, $w_{\rm{dm1}}$, $c^2_{\rm{s,dm}}$ and $\sum m_{\nu}$ for which 95\% CL upper bounds are displayed. The parameters $H_{\rm 0}$  and $\sum m_{\nu}$ are measured in the units of 
km s${}^{-1}$Mpc${}^{-1}$ and $\rm eV$, respectively.}}
\resizebox{\textwidth}{!}{%
\begin{tabular} { c c c c c c}  \hline \hline 
 Parameter &  CMB   & CMB+BAO  & CMB+BAO+PP & CMB+BAO+PP+KiDS-1000 & CMB+BAO+PP+DES\\ 
\hline
$10^{-2}\omega_{\rm b }$ &$2.204^{+0.055}_{-0.056}$& $2.214^{+0.057}_{-0.055} $  & $2.207^{+0.047}_{-0.045} $ & $2.214^{+0.052}_{-0.053} $ &$2.211^{+0.048}_{-0.052} $\\
 
 $\omega_{\rm dm }  $ & $0.1087^{+0.0095}_{-0.0100}$ & $0.1096^{+0.0076}_{-0.0079}$& $0.1095^{+0.0068}_{-0.0064}$ & $0.1109^{+0.0063}_{-0.0067}$& $0.1118^{+0.0069}_{-0.0064}$\\

$100 \theta_{s }  $ & $1.04240^{+0.00100}_{-0.00098}$& $1.04240^{+0.00100}_{-0.00100}$ & $1.04255^{+0.00085}_{-0.00093}$ & $1.04240^{+0.00110}_{-0.00100}$ & $1.04238^{+0.00098}_{-0.00100}$\\

$\ln10^{10}A_{s }$  & $3.023^{+0.039}_{-0.043}$ & $3.028^{+0.036}_{-0.040} $   & $3.023^{+0.038}_{-0.039}$ & $3.027^{+0.040}_{-0.036}$ & $3.031^{+0.038}_{-0.039} $\\

$n_{s } $ &$0.954^{+0.020}_{-0.020}$  & $0.957^{+0.020}_{-0.020} $& $0.955^{+0.018}_{-0.016}$ & $0.957^{+0.019}_{-0.019} $& $0.956^{+0.018}_{-0.019}$\\

$\tau_{\rm reio }   $ & $0.049^{+0.017}_{-0.018}$ &$0.051^{+0.016}_{-0.017}$ & $0.050^{+0.016}_{-0.015} $& $0.051^{+0.017}_{-0.018} $& $0.052^{+0.017}_{-0.017}   $\\
 
$w_{\rm{dm0}}$ & $<3.7\times 10^{-3}$& $<3.3\times 10^{-3}$ & $<2.4\times 10^{-3}$ & $<2.5\times 10^{-3}$ &$<2.2\times 10^{-3}$\\

$w_{\rm{dm1}}$ &$<3.8\times 10^{-3}$ & $<3.3\times 10^{-3}$& $<3.1\times 10^{-3}$& $<2.4\times 10^{-3}$ & $<2.5\times 10^{-3}$\\

 $c^2_{\rm{s,dm}}$ & $<1.3\times 10^{-7}$ & $<1.3\times 10^{-7}$ & $<1.2\times 10^{-7}$ & $<1.1\times 10^{-7}$ &$<0.8\times 10^{-7}$ \\

$\sum m_{\nu} $ & $<0.88$& $<0.86$& $<0.87$ &$<0.69$ & $<0.58$\\

$N_{\rm{eff}}   $ & $2.74^{+0.40}_{-0.40} $ & $2.79^{+0.42}_{-0.39} $& $2.74^{+0.37}_{-0.33}$ & $2.78^{+0.41}_{-0.40}$ &$2.78^{+0.41}_{-0.37}      $\\

$M_B $ & $--$& $--$& $-19.482^{+0.061}_{-0.058} $ &$-19.470^{+0.070}_{-0.071}$ & $-19.470^{+0.067}_{-0.069}$\\

\hline
 
$\Omega_{\rm{m} }$ &$0.305^{+0.053}_{-0.051}$  & $0.304^{+0.018}_{-0.017} $& $0.314^{+0.015}_{-0.015} $& $0.311^{+0.016}_{-0.015} $ &$0.312^{+0.016}_{-0.015}   $\\
  
  $H_{\rm 0}$ & $66.9^{+4.5}_{-4.2}$ & $67.2^{+2.3}_{-2.2} $& $66.1^{+1.9}_{-1.8} $ &$ 66.5^{+2.2}_{-2.3} $& $66.5^{+2.1}_{-2.1}$\\
 
  $\sigma_{8} $ & $0.732^{+0.074}_{-0.067}$ & $0.732^{+0.058}_{-0.052}$  & $0.720^{+0.058}_{-0.049}$ & $0.737^{+0.042}_{-0.039}$ &$0.754^{+0.037}_{-0.036} $\\

  $S_8$& $0.737^{+0.063}_{-0.058}$ & $0.737^{+0.059}_{-0.055}$ & $0.736^{+0.058}_{-0.050}$ & $0.751^{+0.037}_{-0.033} $&$0.768^{+0.032}_{-0.033} $\\

  \hline \hline

\end{tabular}}
\end{table*}

\begin{figure}[!ht]\centering
\includegraphics[width=8.4cm]{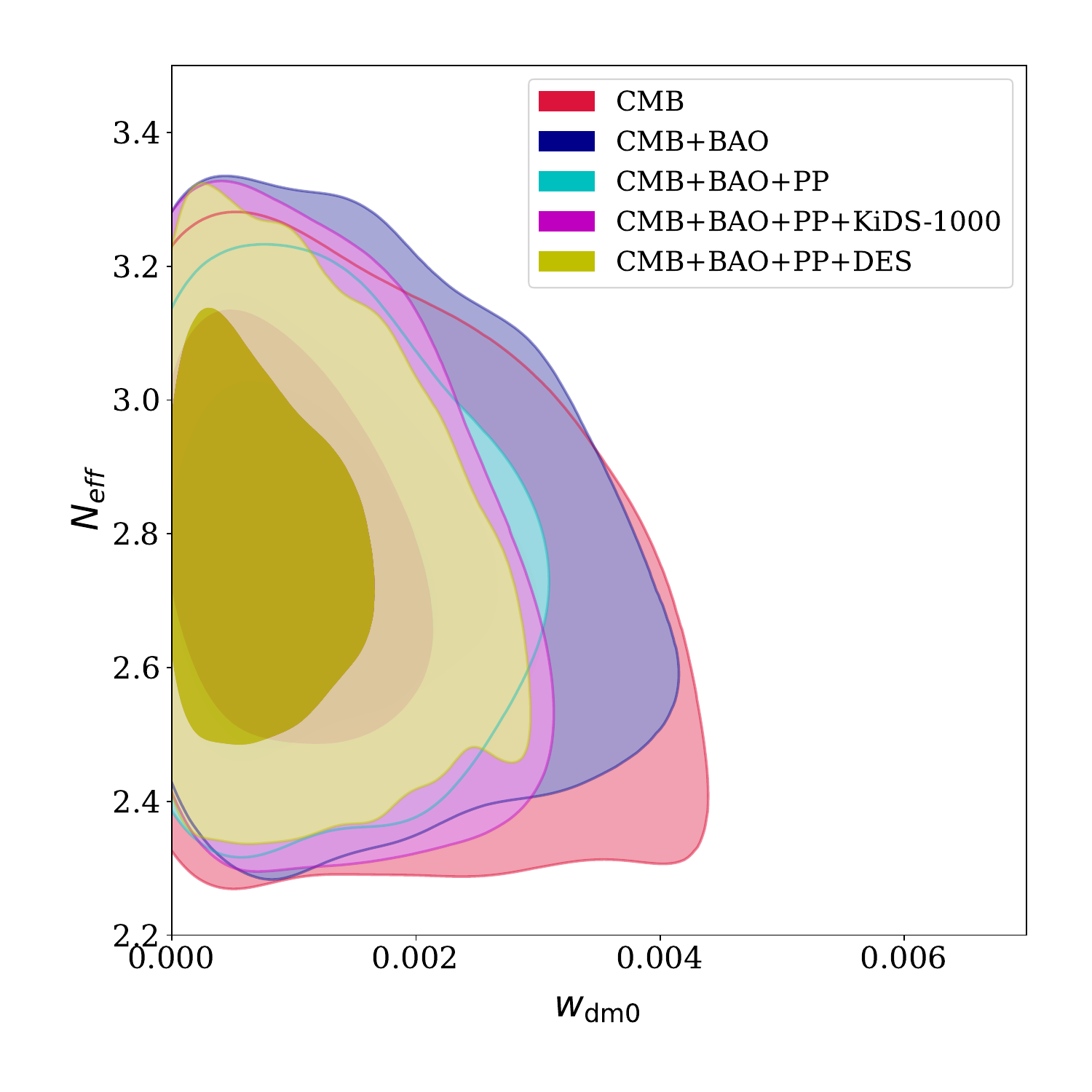} 
 \includegraphics[width=8.4cm]{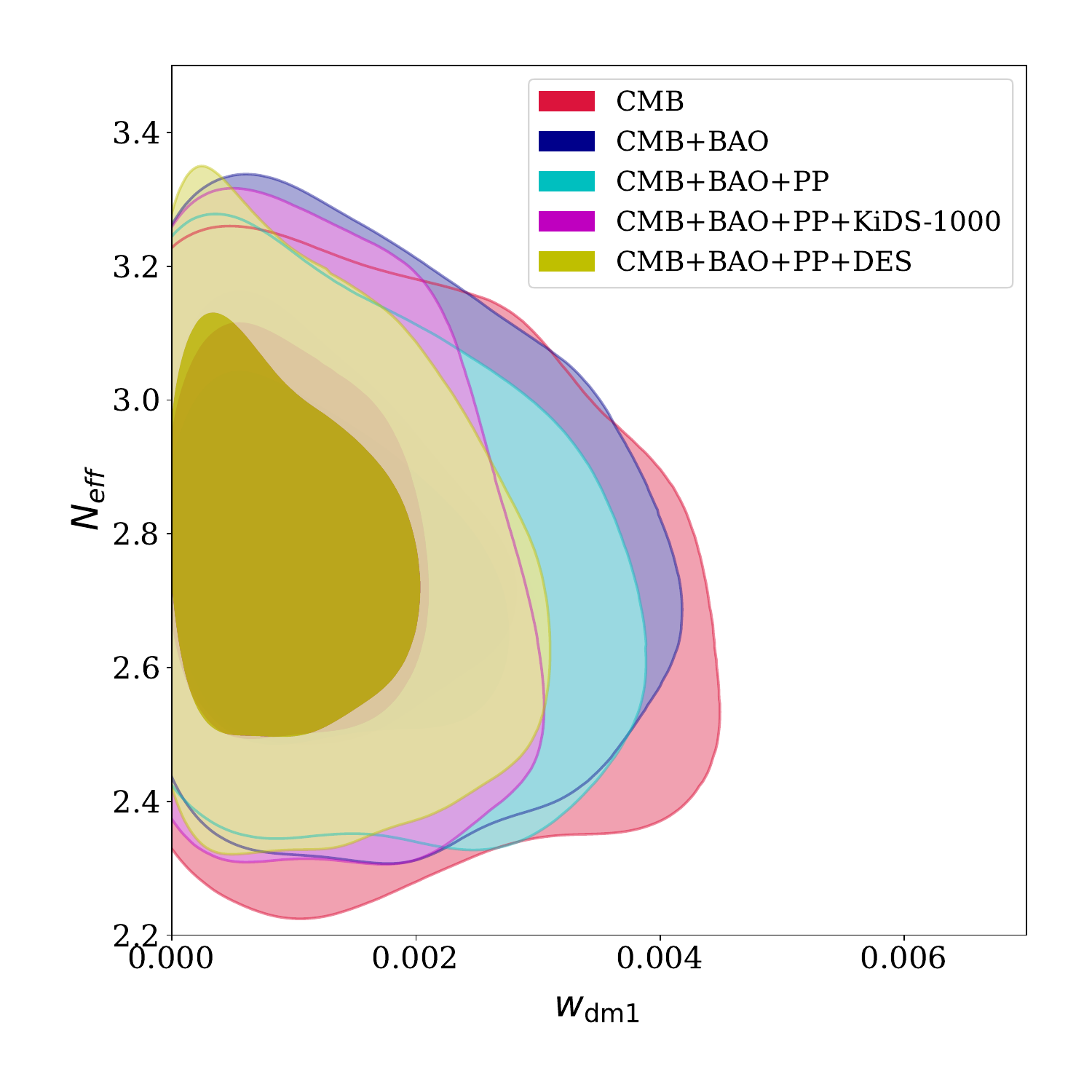} 
\caption{Two-dimensional marginalized distributions (68\% and 95\% CL) of $N_{\rm eff}$ versus DM EoS parameters.} 
\label{neffw0w1}
 \end{figure}

Table \ref{Table_M2} summarizes the observational constraints on the full baseline of the $\Lambda_{\nu}$WDM model for the five different combinations of data sets: CMB, CMB+BAO, CMB+BAO+PP, CMB+BAO+PP+KiDS-1000 and CMB+BAO+PP+DES.
From Table \ref{Table_M2}, we can see that the first six standard parameters are in good agreement with the  $\Lambda$CDM predictions \cite{Planck2018}. Let us first discuss the constraints on generalized DM parameters. We have found that both the DM EoS parameters: $w_{\rm dm0}$ and $w_{\rm dm1}$, are of order $10^{-3}$ at 95\% CL with all data combinations. For the CMB+BAO+PP+KiDS-1000 data combination, these are constrained to less than $2.5\times 10^{-3}$ and $2.4\times 10^{-3}$ at 95\% CL, respectively. Again for the CMB+BAO+PP+DES data combination, these are constrained to less than $2.2\times 10^{-3}$ and $2.5\times 10^{-3}$ at 95\% CL, respectively. The constraints on the sound speed of DM, $\hat{c}^2_{\rm s,dm}$  are of the order $10^{-7}$ at 95\% CL for all the data combinations. The most tight constraint is obtained from CMB+BAO+PP+DES data combination: $\hat{c}^2_{\rm s,dm}<0.8 \times 10^{-7}$ at 95\% CL. 
We have not observed any correlation of DM EoS parameters with the sum of neutrino masses, $\sum m_{\nu}$. Thus, the constraints on DM EoS parameters are not affected by varying the neutrino mass.  Although, the DM EoS parameters are affected by varying $N_{\rm eff}$ as they found a slight negative correlation with $N_{\rm eff}$, see Figure \ref{neffw0w1}.  

\begin{figure*}[hbt!]
 \includegraphics[width=8.4cm]{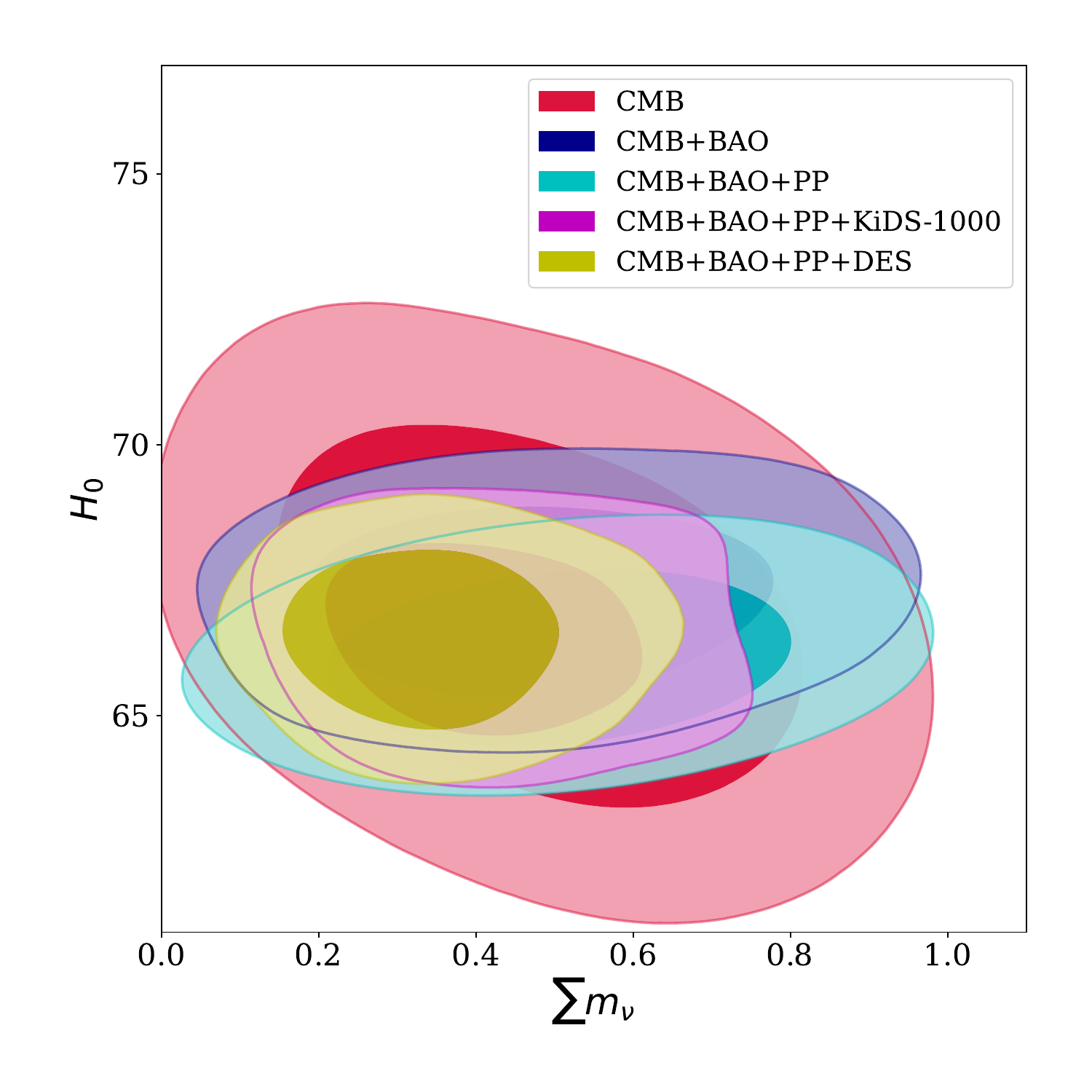} 
 \includegraphics[width=8.4cm]{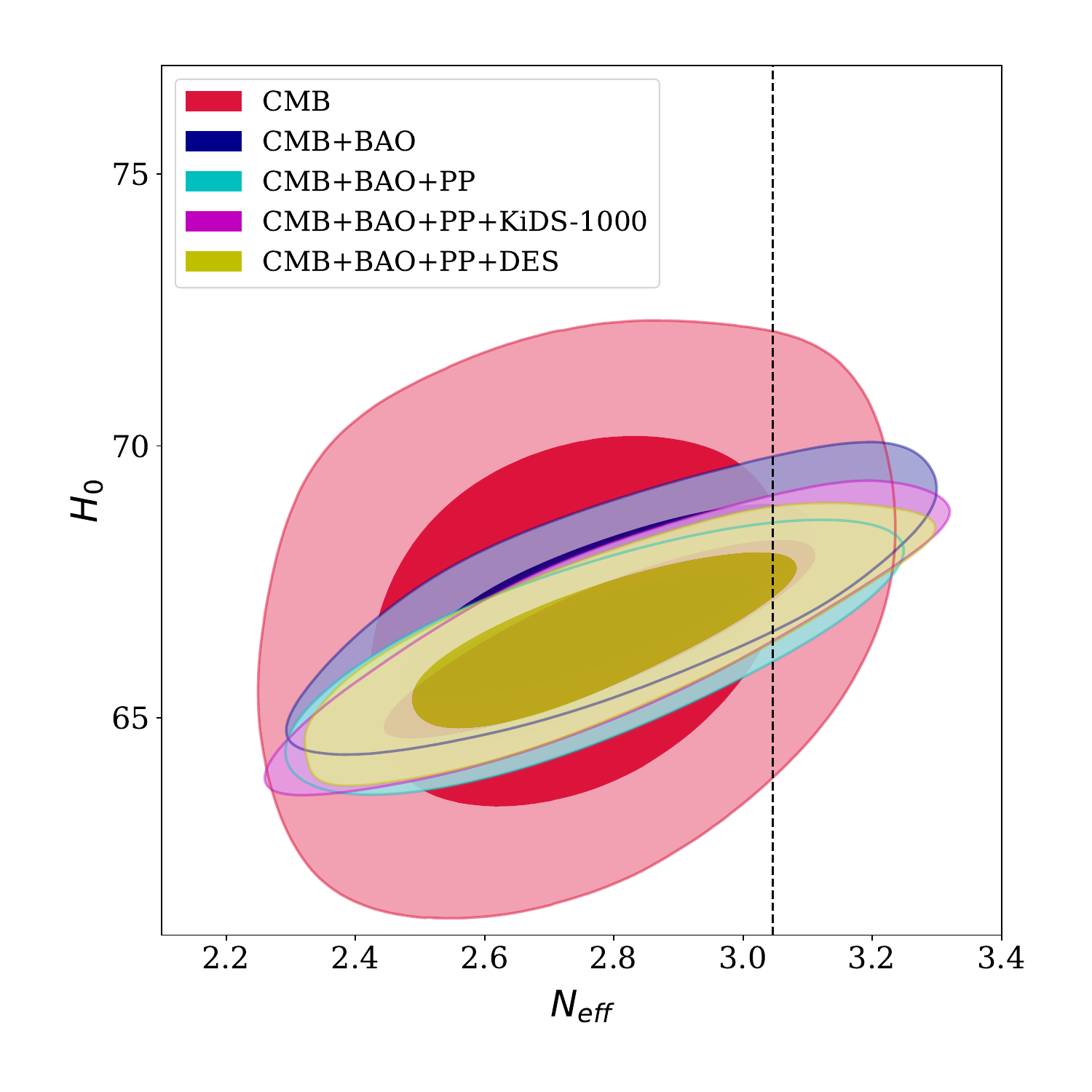} 
 \caption{Two-dimensional marginalized distributions (68\% and 95\% CL) in the plane $\sum m_\nu-H_0$ (left panel) and $N_{\rm eff}-H_0$ (right panel). In the right panel, the dotted vertical line relates to $N_{\rm eff} \approx 3.046$ predicted from $\Lambda$CDM  model \cite{Planck2015}.} 
 \label{h0_mnuneff}
  \end{figure*}

We find significantly tight upper bounds on the sum of neutrino masses in all cases. The most tight upper bound, $\sum m_{\nu}<0.58$ eV at 95\% CL is obtained from the CMB+BAO+PP+DES data combination.  With the other data combinations, we have obtained relatively weaker bounds.  It is known that the observational upper bounds on neutrino masses become weaker when analyzed in the context of extended cosmological scenarios \cite{Choudhury2,Valentino,Giusarma13,Yang2, kumar22}. Thus, in the present analysis, we have found slightly weaker constraints on $\sum m_{\nu}$ compared to  the bounds obtained from final Planck release which yields, $\sum m_\nu < 0.13$ eV at 95\% CL from CMB+BAO, based on $\Lambda$CDM model \cite{Planck2018}.  The parameter $\sum m_\nu$ shows a negative correlation with $H_0$ in case of CMB data, (see the left panel of Figure \ref{h0_mnuneff}) whereas the parameter $N_{\rm eff}$ exhibits a positive correlation with $H_0$, (see the right panel of Figure \ref{h0_mnuneff}). These constraints on $N_{\rm eff}$ are relatively weaker but in agreement with the constraints from final Planck results \cite{Planck2018}. Likewise, the constraints on $H_0$ in all the cases are relatively weaker but consistent with $H_0 = 67.27\pm 0.60{\rm \,km\, s^{-1}\, Mpc^{-1}}$ (68\% CL)~\cite{Planck2018} inferred from \textit{Planck} 2018 assuming the standard $\Lambda$CDM model. It means the $H_0$ tension is not relaxed in the $\Lambda_{\nu}$WDM model.

\begin{figure*}[hbt!]
 \includegraphics[width=8.4cm]{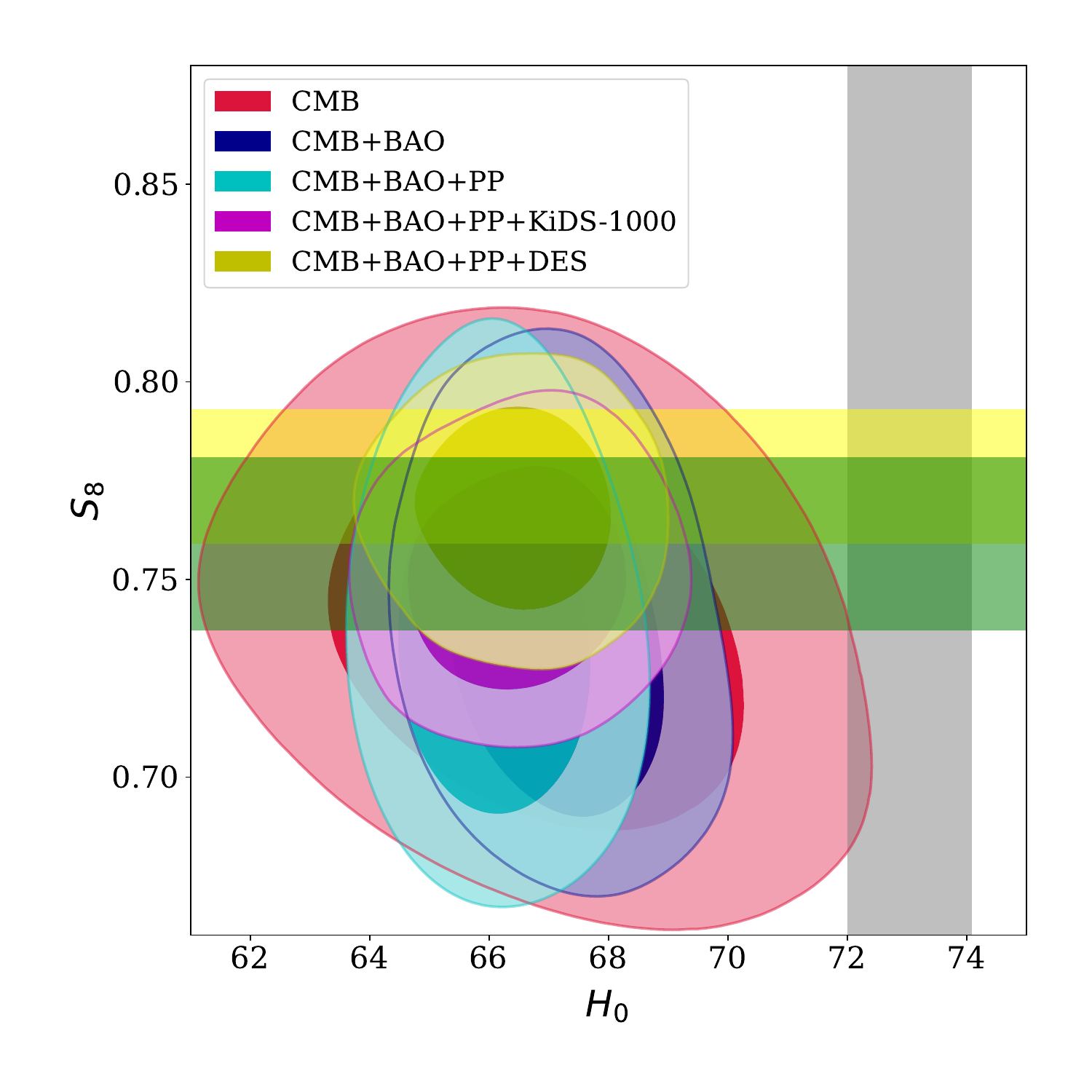} 
 \includegraphics[width=8.4cm]{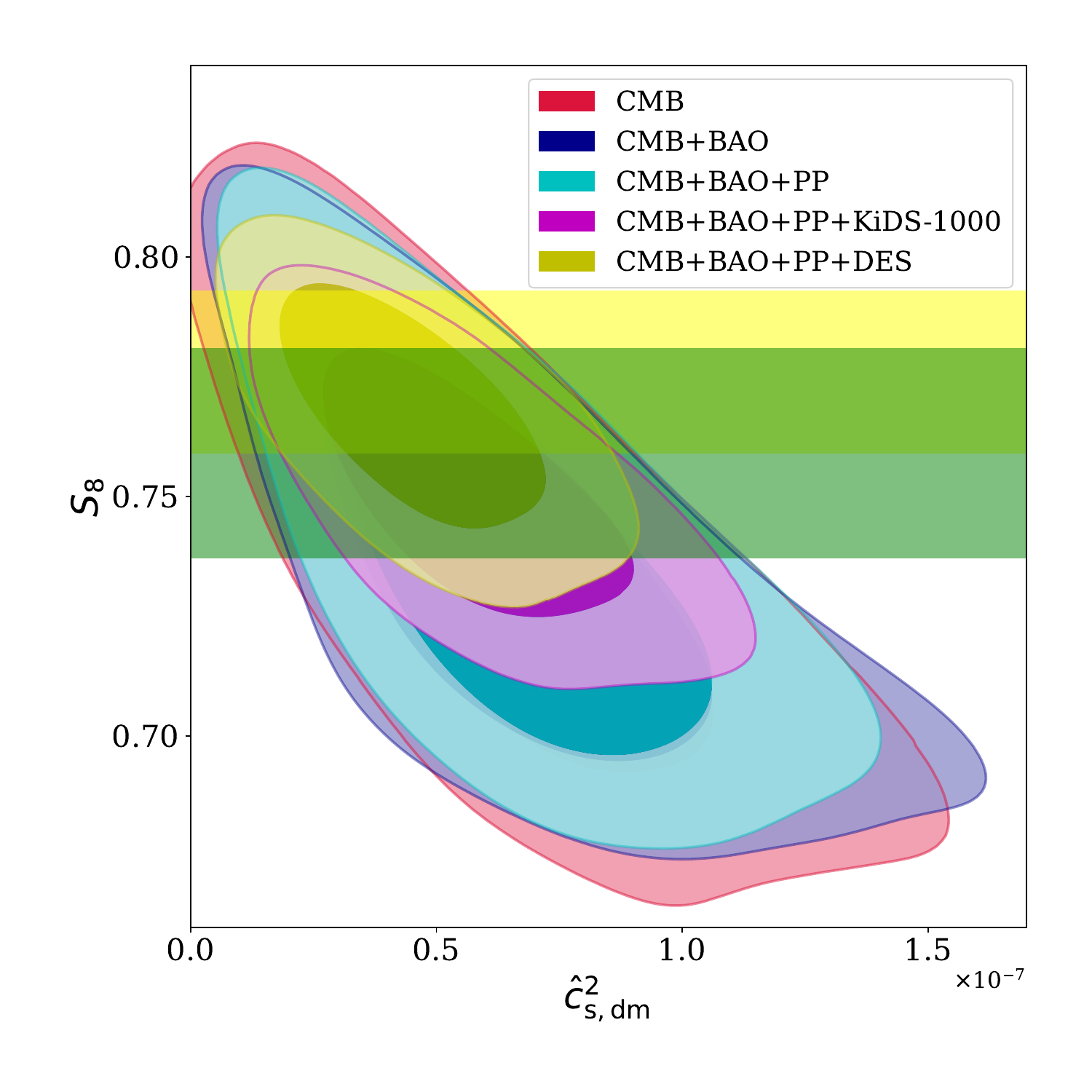} 
 \caption{Two-dimensional marginalized distributions (68\% and 95\% CL) in the plane  $H_0$-$S_8$(left panel) and $\hat{c}^2_{\rm s,dm}-S_8$ (right panel) from the different data combinations. The vertical grey band in the left panel  stands for $H_0 = 73.04 \pm 1.04$ km s$^{-1}$Mpc$^{-1}$ \cite{Riess:2021jrx} whereas the horizontal yellow and green bands in both panels respectively correspond to $S_8$ obtained by DES ($S_8=0.776 \pm 0.017$, \cite{DES:2021wwk}) and KiDS-1000 ($S_8=0.759 \pm 0.022$, \cite{KiDS:2020suj}).} 
 \label{fig3}
  \end{figure*}
It is interesting to observe lower mean values of $S_8$ in $\Lambda_{\nu}$WDM model in all the analyses compared to $S_8=0.834 \pm 0.016$ inferred from \textit{Planck} 2018 ~\cite{Planck2018} $\Lambda$CDM model. Figure \ref{fig3} (left panel) shows the parametric space in the plane $H_0$ - $S_8$ with all the data combinations where one can see that the allowed range of $S_8$ (horizontal bands) from LSS measurements  pass through the central region of the contours.  From Figure. \ref{fig3} (right panel), we can see that the parameter $S_8$ is in a perfect negative correlation with sound speed of DM, $\hat{c}^2_{\rm s,dm}$. The reason is that the presence of sound speed sufficiently reduces the growth of matter density fluctuations on the scale below the diffusion length \cite{gdm01}. Thus, we have obtained lower mean values of $S_8$ but with large errors due to the strong degeneracy with $\hat{c}^2_{\rm s,dm}$. Thus, without loss of generality, we can conclude that the $S_8$ tension is alleviated at $68\%$ CL within $\Lambda_\nu$WDM model.


\subsection{Bayesian model comparison}
It is worthwhile to note that a statistical comparison of the derived model with reference model exhibits the consequences of the considered model. Therefore, we use the Akaike Information Criteria (AIC) \cite{AIC01, AIC02}, defined as 
 \begin{equation} \nonumber
 \text{AIC} = -2 \ln  \mathcal{L}_{\rm max} + 2N \quad = \chi_{\rm min}^2 + 2N,
 \end{equation}
where $ \mathcal{L}_{\rm max}$ and $N$ denote the maximum likelihood function and the total number of free parameters in the model baseline. To compare the derived model $i$ with a well known best fit reference model $j$, one may compute the AIC difference between two models, i.e.,  $\Delta\text{AIC}_{ij} =  \text{AIC}_{i}- \text{AIC}_{j}$. This difference may play an important role in quantify the model $i$ and model $j$. In Ref. \cite{AIC05}, the authors have investigated that one model is better than other if the AIC difference between these models is greater than a threshold value $\Delta_{\rm threshold}$. According to the thumb rule of AIC, we note that $\Delta_{\rm threshold} = 5$ is a universal threshold value which is used as a property of the model for comparison. In Liddle \cite{Liddle:2007}, it is clearly mentioned that one model is better compared to the other model depending upon the minimum threshold AIC difference. 
\begin{table}[h]\centering
\caption{\label{evidence}{}Difference of AIC values  of $\Lambda_\nu$WDM model with respect to $\Lambda$CDM model with five data combinations.}
\begin{tabular}{l c c }
\hline \hline
Data  &   $\Delta \rm AIC_{\Lambda_{\nu}WDM}$ \\
\hline
CMB   & 6.06  \\
CMB + BAO  &   11.42 \\
CMB + BAO + PP  &  3.00\\
CMB + BAO + PP + KiDS-1000  &  $-6.68$ \\
CMB + BAO + PP + DES &     $-5.42$\\

\hline \hline
\end{tabular}
 \end{table}
 
Table \ref{evidence} summarizes the $\Delta \rm AIC$ values of the $\Lambda_{\nu}$WDM model for all the data combinations. We see that with the CMB and CMB+BAO data combinations, the standard $\Lambda$CDM model is strongly preferred over $\Lambda_{\nu}$WDM model; with the CMB+BAO+PP data set the difference between the models is not statistically significant, whereas with both CMB+BAO+PP+KiDS-1000 and CMB+BAO+PP+DES data combinations, the $\Lambda_{\nu}$WDM model is favored over the  $\Lambda$CDM model.  It is important to note that in addition to the free parameters due to extended properties of DM and neutrinos, here in $\Lambda_{\nu}$WDM model, we have additional parameters. Also, the CMB and LSS data sets  are  in tension within the framework of $\Lambda$CDM model. Thus, it is not surprising to observe possible evidence for some new physics beyond the $\Lambda$CDM  via the Bayesian model comparison in the presence of such data sets. On the other hand, the possible existence of systematic effects on these data is not yet well determined, possibly causing the tension with $\Lambda$CDM. Therefore, the use of these data is important to study the extended models of $\Lambda$CDM, such as $\Lambda_{\nu}$WDM model investigated in this work.
\section{Concluding remarks} \label{CR}
In the present work, we have investigated an extension of $\Lambda$CDM model by considering the generalized DM properties: a possible time dependence of the equation of state (EoS) of DM via CPL parameterization, and the non-null sound speed taken as constant plus the neutrino properties. We have derived robust constraints on the generalized DM parameters, neutrino properties, and analyzed their possible correlations with the other parameters. With all data combinations considered in this work, the DM parameters $w_{\rm dm0}$,  $w_{\rm dm1}$ are constrained up to the order $10^{-3}$, and $\hat{c}^2_{\rm s,dm}$ is constrained up to the order $10^{-7}$ all at 95\% CL. We notice that the extended DM parameters are very close to null values with no significant evidence beyond the standard CDM paradigm, leading to the conclusion that the present observational data favor DM as a pressureless fluid. 
The present analysis yields significantly tight upper bounds on the sum of neutrino masses in all cases.   Also, we have observed that the neutrino properties exhibit correlation with DM extended parameters and also with other model parameters.
It is worthy to mention that we have found significantly lower mean values of $S_8$ in all cases, which are in good agreement with the LSS measurements (see Figure \ref{fig3}). Thus, the well known $S_8$ tension is reconciled within the $\Lambda_\nu$WDM model.

\section*{Acknowledgment}
The authors gratefully thank the reviewer for fruitful and constructive comments/suggestions to enhance the quality of the paper. 


\end{document}